\documentclass[preprint,12pt]{elsarticle}

\usepackage{graphicx}
\usepackage{bm,amsmath,amssymb}

\newcounter{bla}

\journal{Computer Physics Communications}

\begin{document}

\begin{frontmatter}

%% Title, authors and addresses

%% use the tnoteref command within \title for footnotes;
%% use the tnotetext command for the associated footnote;
%% use the fnref command within \author or \address for footnotes;
%% use the fntext command for the associated footnote;
%% use the corref command within \author for corresponding author footnotes;
%% use the cortext command for the associated footnote;
%% use the ead command for the email address,
%% and the form \ead[url] for the home page:
%%
%% \title{Title\tnoteref{label1}}
%% \tnotetext[label1]{}
%% \author{Name\corref{cor1}\fnref{label2}}
%% \ead{email address}
%% \ead[url]{home page}
%% \fntext[label2]{}
%% \cortext[cor1]{}
%% \address{Address\fnref{label3}}
%% \fntext[label3]{}

\title{SPOCK*: A simple program for simulating knotted and concatenated polymer rings off-lattice}

%% use optional labels to link authors explicitly to addresses:
%% \author[label1,label2]{<author name>}
%% \address[label1]{<address>}
%% \address[label2]{<address>}

\author[a]{Franco Ferrari\corref{author}}
\author[a]{Marcin R. Pi\c{a}tek}
%\author[b]{Third Author}

\cortext[author] {Corresponding author.\\\textit{E-mail address:} franco.ferrari@usz.edu.pl}
\address[a]{CASA* and Institute of Physics, University of Szczecin,
  Szczecin, Poland}

\begin{abstract}
%% Text of abstract
The purpose of this work is to present SPOCK*, a Monte Carlo code
specifically written to investigate the thermodynamic and mechanical properties of polymers in the presence of topological constraints.
The interactions between the monomers are described by a Lennard-Jones potential.
Pulling forces can be applied to one or more monomers. 
Simple and fast algorithms have been implemented to preserve the topology and to compute the energy of the sampled conformations.
After a new conformation is accepted, only the difference of energy between the new and the old conformations needs to be evaluated. In this way the simulation time grows linearly with the polymer size. A strategy based on the fluctuations of the specific heat capacity has been developed in order to avoid bottlenecks like the trapping of the system in a deep local minima at low temperature.
Currently, the averages of the following observables are computed: specific heat capacity, elongation and gyration radius.
\\

% All CPiP articles must contain the following
% PROGRAM SUMMARY.

%\noindent \textbf{PROGRAM SUMMARY/NEW VERSION PROGRAM SUMMARY}
  %Delete as appropriate.

   \end{abstract}
\end{frontmatter}

%% main text
\section{Introduction}
\label{introd}
At present there are several software tools to simulate polymer systems, like for instance
LAMMPS\cite{lammps}, GROMACS\cite{gromacs}, HOOMD-blue\cite{hoomd} and
DL\verb|_|POLY\cite{dlpoly}.
However, there are no publicly available programs that specifically deal with the thermodynamic and mechanical properties of polymers in the presence of topological constraints.
The aim of the present article is to introduce SPOCK*, a simple Monte Carlo code
to compute the average of the observables of a system of concatenated knotted polymer rings.
SPOCK* is the acronym for Simple Program for Observables' Computation for Knots, the \verb|*| indicating other kinds of topological soft matter like for instance concatenated knotted rings.
Polymers are modeled as a system of beads connected together by bonds to form rings that can additionally  be knotted and/or concatenated together. The Monte Carlo simulations are  performed off-lattice.
The monomer-monomer interactions are taken into account by a Lennard-Jones potential. A constant pulling force can be applied to one or more monomers.
The expectation values of the following observables are currently implemented:
specific heat capacity, square gyration radius, elongation following the application of a force on a given monomer and energy. For each quantity
the related standard deviation from its average value is calculated.
Sampling starts from a seed conformation of the polymer system. Further conformations are obtained like in Ref.~\cite{baumgartner} by
rotating randomly selected monomers around a circle by a random angle.
Topology breakings are prevented by checking that in the portion of space interested by the rotation there are no other components of the system apart from the  monomer to be rotated and the bonds connecting it to the neighboring monomers.
New conformations are selected acconrding to Metropolis Monte Carlo algorithm.

The code has been designed for simplicity and speed.
A boost in speed has been achieved by rearranging all calculations in such a way that the simulation time scales linearly with the total number of monomers. This goal has been realized by computing after each random transformation only the difference of energy between the new and the old conformations. In this way, the computation of the energy of the system requires only a single sum instead of the double sum that it is usually necessary to perform in the case of a two-body potential like that of Lennard-Jones. Also a strategy for a fast equilibration of the system has been implemented. Especially for long polymers and at low temperatures, the need of equilibration after starting from a seed conformation that can be out of equilibrium could slow down simulations considerably and the system could get trapped in a local minimum. The efficiency of the adopted strategy has been verified in the case of long polycatenanes containing a large number (more than thousand) of monomers.
The algorithm for preserving the topology of the system after a new conformation is sampled is based on geometry.
The time
for checking potential 
topology breakings grows linearly with the polymer size.

%% The Appendices part is started with the command \appendix;
%% appendix sections are then done as normal sections
%% \appendix

\section{Methodology}
 \label{method}
\subsection{The sampling algorithm}
The code is based on the Metropolis Monte Carlo algorithm and is written in Fortran~95.  It may be compiled using the GNU gfortran compiler using for instance the command:
\verb|gfortran-15 -Ofast -flto=auto *.f90|.
Further machine-dependent options may be added depending on the available hardware thanks to the option \verb|-march|.

The sampling procedure starts from a seed consisting of \verb|NKNOTS| rings of the same length \verb|L| that are knotted and possibly concatenated with themselves. An extensive library of seeds is available. A separate program is provided that is able to increase the size of the seed by an integer factor.
The seed is stored in the \verb|txt| file \verb |seed| written  with the following format:
\begin{center}
\begin{tabular}{ccc}
  $x_{1,1}$&$y_{1,1}$ &$z_{1,1}$\\
  \multicolumn{3}{c}{$\ddots$}\\
  $x_{1,L} $&$y_{1,L}$ &$z_{1,L}$\\
  $x_{1,1} $ &$y_{1,1}$ &$z_{1,1}$\\
    \multicolumn{3}{c}{$\vdots$}\\
  $x_{NKNOTS,1}$&$y_{NKNOTS,1} $&$z_{NKNOTS,1}$\\
  \multicolumn{3}{c}{$\ddots$}\\
  $x_{NKNOTS,L} $&$y_{NKNOTS,L} $&$z_{NKNOTS,L}$\\
  $x_{NKNOTS,1}  $&$y_{NKNOTS,1}$ &$z_{NKNOTS,1}$\\
\end{tabular}
\end{center}
where $x_{KNN,I},y_{KNN,I},z_{KNN,I}$ are the coordinates of the $I-$th monomer of the $KNN-$th ring. In the code, knots are labeled by the integer indices \verb|KNN,KNN1,KNN2...=1,..,NKNOTS|, while latin letters \verb|I,J,...=1,..,L+1| are used for the monomers. An exception is the monomer to which the force is applied, which is denoted by the integer \verb|PAF|.
Moreover, the coordinates of the last conformation $\cal X$ accepted during the sampling process and those of the trial conformation ${\cal X}'$
are respectively denoted by
\begin{center}
\verb|REPX(KNN,I),REPY(KNN,I),REPZ(KNN,I)|\\ and\\
    \verb|REPNX(KNN,I),REPNY(KNN,I),REPNZ(KNN,I)|
    \end{center}
Note that for each knot \verb|KNN=1,..,NKNOTS|, the monomer \verb|L+1| coincides
with the first monomer in order to preserve the continuity of the ring. For instance, \verb|REPX(KNN,L+1)=REPX(KNN,1)|, \verb|REPY(KNN,L+1)=REPY(KNN,1)| and
\verb|REPZ(KNN,L+1)=REPZ(KNN,1)|.

Let $H({\cal X})$ be the energy of the conformation ${\cal X}$. At the moment, in the code the potential energy of the interactions between the monomers consists of
a Lennard-Jones (LJ) potential $V_{LJ}(r)$ with the addition of a  potential $V_F(K,\boldsymbol r;K',\boldsymbol r')$ related to a constant pulling force:
\begin{equation}
  V_{LJ}(r)=4\epsilon\left[
\left(\frac{\sigma}{r}\right)^{12}-\left(\frac{\sigma}{r}\right)^{6}
  \right]
\end{equation}
\begin{equation}
  V_{F}(K,I;K',I')=
\boldsymbol F\cdot(\boldsymbol r_{K,I}-\boldsymbol r_{K',I'})
\end{equation}
Here $r$ is the distance between two monomers, while $\epsilon$ and $\sigma$ are constant parameters. In the code, these parameters are \verb|EPS| and \verb|SIGMA| respectively.  Pulling is realized by keeping fixed the monomer \verb|I'| on the knotted ring \verb|KNN'| and applying a force on the monomer \verb|PAF| of knotted ring \verb|KNNZ|.
By default, \verb|KNN'=1| and  \verb|I'=1|, \verb|KNNZ=NKNOTS/2| and \verb|PAF=L/2|.
Moreover, the pulling force \verb|FORCEZ| is directed along the $z-$axis, for instance 
\verb|FORCEZ=10.0|.
It is easy to change this setup adding an arbitrary number of forces acting on different monomers and  pointing along any direction.
 It is also possible to add to the LJ-potential a Coulomb potential or other potentials, provided they are two-body potentials.

Following the Metropolis algorithm, the trial conformation with energy $H({\cal X}')$ is accepted with probability
\begin{equation}
  p({\cal X}|{\cal X}')=\min\left(1,e^{-\frac{(H({\cal X}')-H({\cal X}))}{T}}\right)\label{metropolis}
 \end{equation}
  $H$ and
$T$ are rescaled and dimensionless. In particular, the temperature $T$ is given in thermodynamics units in which the Boltzmann constant $k_B$ is equal to one.
\begin{figure}[h]
  \begin{center}
    \includegraphics[height=3cm]{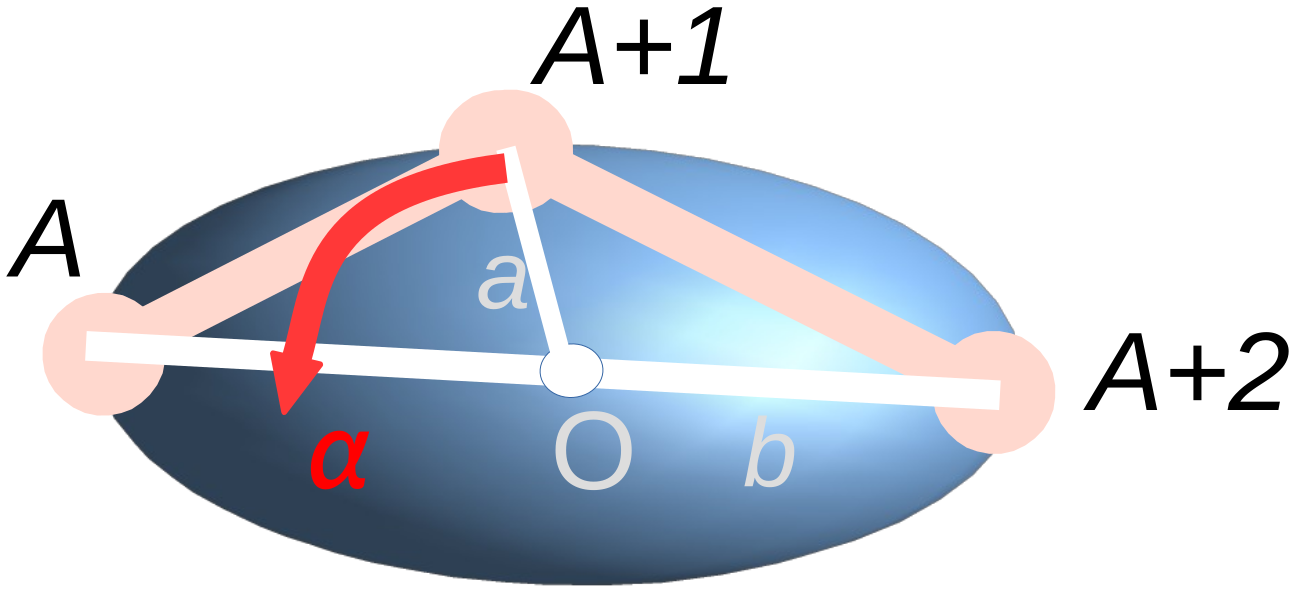}
  \end{center}
  \caption{a}
  \label{fig1}
\end{figure}
\subsection{The random moves and enforcement of the topological constraints}
To perform the sampling, starting from the initial seed ${\cal X}^{(0)}$, the $\nu-th$ conformation ${{\cal X}^{(\nu+1)}}$  is obtained from ${\cal X}^{(\nu)}$ by applying
random transformations to a portion of ${\cal X}^{(\nu)}$ consisting of \verb|NPOINTS| consecutive monomers \verb|A|,\verb|A+1|,...,\verb|A+NPOINTS|, where \verb|A=1,...,L|. The first monomer \verb|A| and the knot $K$ in which the part to be changed is located, are randomly chosen.
The applied transformations are rotations around a circle of the beads representing the monomers like in Ref.~\cite{baumgartner}. Namely, each bead \verb|A+1|,$\ldots$,\verb|A+NPOINTS|
is rotated by random angles $\alpha_1,\ldots,\alpha_{A+NPOINTS}$, see Fig.~\ref{fig1}.

After each rotation, a check of possible topology breakings is made. A newly generated conformation  ${{\cal X}^{(\nu+1)}}$ is rejected whenever the topological state of the previous conformation is not preserved.
To preserve the topology, an algorithm based on geometry is provided. Looking at Fig.~\ref{fig1}, it is possible to see that, however the point \verb|A+1| is rotated on a circle,  the segments
$|A(A+1)|$ and $|(A+1)(A+2)|$
 connecting respectively the monomers $A,A+1$ and $A+1,A+2$,
rotate inside an ellipsoid of semiaxes $a=|\boldsymbol r_{K,A+1}-\boldsymbol O|$ and $b=|\boldsymbol r_{K,A+2}-\boldsymbol r_{K,A+1}|$.
Here $\boldsymbol O=(O_x,O_y,O_z)$ is the radius vector identifying the point $O$ in space that is the projection of monomer $A+1$ on the segment $\boldsymbol r_{K,A+2}-\boldsymbol r_{K,A+1}$, see Fig.~\ref{fig1}.
In formulas:
\begin{equation}
  \boldsymbol O=\boldsymbol r_{K,A}+(\boldsymbol r_{K,A+2}-\boldsymbol r_{K,A})
  \frac{(\boldsymbol r_{K,A+1}-\boldsymbol r_{K,A})\cdot(\boldsymbol r_{K,A+2}-\boldsymbol r_{K,A})}{|\boldsymbol r_{K,A+2}-\boldsymbol r_{K,A} |^2}
\end{equation}
The ellipsoid satisfies the equation:
 \begin{equation}
\frac{(x-O_x)^2}{b^2}+\frac{[(y-O_y)^2+(z-O_z)^2]}{a^2}=1\label{ellipsoid}
 \end{equation}
 Clearly, the topology of the new conformation ${\cal X}^{(\nu+1)}$
 obtained after a random rotation of the monomer \verb|A+1| belonging to the knotted ring $K$
does not change with respect to that of the old conformation ${\cal X}^{(\nu)}$
if inside the ellipsoid defined by Eq.~(\ref{ellipsoid}) there are no segments belonging to the polymer system
except the two segments $\boldsymbol r_{K,A+1}-\boldsymbol r_{K,A}$ and $\boldsymbol r_{K,A+2}-\boldsymbol r_{K,A+1}$
 that are involved in the rotation.
As a consequence, to assess if the topology is destroyed or not after a random transformation, all segments  $\boldsymbol r_{K',I+1}-\boldsymbol r_{K',I}$ are checked for $K'=1,\ldots, NKNOTS$ and
$I=1,\ldots,L$, with the further restriction that $I\neq A,A+1,A+2$ if $K'=K$.
The points $(x_{K',I}(t),y_{K',I}(t),z_{K',I}(t))$ belonging to one of these
 segments are parametrized as follows:
 \begin{equation}
   (x_{K',I}(t),y_{K',I}(t),z_{K',I}(t))=-\boldsymbol r_{K',I}+t(-\boldsymbol r_{K',I+1}-\boldsymbol r_{K',I})\label{segment}
 \end{equation}
where $t$ is a parameter such that $0\le t\le 1$.
 If a point of  the segment $\boldsymbol r_{K',I+1}-\boldsymbol r_{K',I}$ is inside the ellipsoid (\ref{ellipsoid}), then there should be
 a solution of the equation
\begin{equation} \frac{(x_{K',I}(t)-O_x)^2}{b^2}+\frac{[(y_{K',I}(t)-O_y)^2+(z_{K',I}(t)=O_z)^2]}{a^2}=r^2
  \label{tpc}\end{equation}
 for some real value of $t$ in the interval $0\le t\le 1$. Eq.~(\ref{tpc}) is a quadratic equation in $t$ which admits two solutions $t_1$ and $t_2$.
 The existence or not of such solutions depends on the discriminant $\Delta$ of that quadratic equation. $\Delta$ can be easily calculated.
 To decide if a given segment $\boldsymbol r_{K',I+1}-\boldsymbol r_{K',I}$
 is at least partly inside the ellipsoid defined by Eq.~(\ref{ellipsoid}) or not, the following criteria are applied:
 \begin{enumerate}
 \item[Case 1:] $\Delta<-\epsilon$, where $\epsilon$ is a small and positive constant that is introduced in order to take into account possible rounding errors in the calculations.
  A cautious choice is  $0.00001\le\epsilon\le 0.001$.
  In case 1, the solutions $t_1$ and $t_2$ of the quadratic equation (\ref{tpc}) are complex.
As a consequence, the segment $\boldsymbol r_{K',I+1}-\boldsymbol r_{K',I}$
does not penetrate the surface of the ellipsoid inside which the random rotation leading to the new conformation ${\cal X}^{(\nu+1)}$ has taken place.
The flag \verb|FLAG1| is put equal to zero and the check is extended to the next segment.
 \item[Case 2:] $-\epsilon\le \Delta\le \epsilon$. In this case $\Delta$ is small and the two solutions $t_1,t_2$ approximately coincide, i. e. $t_1\sim t_2$.
   This means that the segment  $\boldsymbol r_{K',I+1}-\boldsymbol r_{K',I}$ is approximately tangent to the ellipsoid defined by Eq.~(\ref{ellipsoid}). Depending if
   the segment  $\boldsymbol r_{K',I+1}-\boldsymbol r_{K',I}$ slightly penetrates the volume if the ellipsoid or not, a  topology breaking may potentially occur.  To simplify the calculations, if case 2 is verified the flag \verb|FLAG1| is put equal to \verb|1|, implying that the check of possible topology violations by the remaining segments is stopped. The new conformation ${\cal X}^{(\nu+1)}$ is rejected and a new one is generated. 
   The reason for assuming that the topology is violated if $\Delta$ is in
the interval $[-\epsilon, \epsilon]$
   is that the solutions $t_1$ and $t_2$ of a quadratic equation of the type $at^2+2bt+c=0$ are given by $t_{1,2}=\frac{-b\pm\sqrt{\Delta}}{a}$, where
   $\Delta= ac-b^2$. Even if $\Delta$ is small, there could be particular situations in which $a$ is small too, so that the ratio 
$\frac {\sqrt{\Delta}}{a}$ becomes not negligible. To analyse all possibilities during the sampling would be time consuming.
   This choice is not a great limitation owing the fact that the parameter $\epsilon$ is small. 
 \item[Case 3:] $\Delta> \epsilon$
   In this case Eq.~(\ref{tpc}) admits two real solutions $t_1\ne t_2$.
   If  one of the following two conditions occurs:
   \begin{enumerate}
   \item $t_1\in[0,1]$;
     \item  $t_2\in [0,1]$,
   \end{enumerate}
   this means that the segment penetrates the ellipsoid and the topology is thus not preserved. As a consequence, the flag \verb|FLAG1| is put equal to \verb|1|, the check of the remaining segments is stopped and the new conformation  ${\cal X}^{(\nu+1)}$ is  discarded.
 \end{enumerate}
 If the flag \verb|FLAG1|is still equal to zero after all segments have been checked, then the new conformation ${\cal X}^{(\nu+1)}$  is accepted.
 The subroutines implementing in the code the above algorithm for topology preservation are
 \verb|NROTATE|, \verb|TOP_CHECK| and \verb|TOP_CHECKR|. The subroutine \verb|NROTATE| performs random rotations of the set of \verb|NPOINTS| monomers \verb|A+1|,...,\verb|A+NPOINTS| on the knotted ring $K$. After  one of these points has been rotated, the potential topology breakings due to segments of $K$ are checked by subroutine \verb|TOP_CHECK|. Potential breakings of the topological relations due to the segments of all other knotted rings $K'\ne K$ are taken into account by subroutine  \verb|TOP_CHECKR|. When running codes with only one knot, the call to  \verb|TOP_CHECKR| in the file \verb|nrotate.f90| should be commented.
\subsection{Energy calculation and "cooling" strategy}
While the calculation of the contribution $E_F$ of  the pulling force $\boldsymbol F$ to the total energy of the system is simple and very fast, this is not the case of the interactions between the monomers. For this, the evaluation of a double sum over all monomers of the knotted polymers is  necessary:
\begin{equation}
  E_{LJ}=\frac 12\sum_{K=1}^{NKNOTS} \sum_{I\ne J=1}^L
  V_{LJ}(\boldsymbol r_{K,I} - \boldsymbol r_{K,J})
  +\frac 12
  \sum_{K\ne K'=1}^{NKNOTS}{\sum_{{I,J=1}}^{ L}} V_{LJ}(\boldsymbol r_{K,I} - \boldsymbol r_{K',J})\label{LJenergy}
\end{equation}
The  above energy calculation for each sampled conformation requires a time $\tau$ that scales as $\tau\propto(NKNOTS\cdot L)^2$. Clearly, when the number
 \verb|NKNOTS| of knotted polymer rings and the number  \verb|L| of monomers in each ring grow larger and larger, the simulations become increasingly slow.
 To overcome this problem, the total potential energy $E_{LJ}$ due to the monomer-monomer interactions is computed only once at the beginning of the simulation by the subroutines \verb|ENERGYSYS| and \verb|ENERGYR|. \verb|ENERGYSYS| is used to evaluate the potential energy of the monomers in a single knot, see first term in the right hand side of Eq.~(\ref{LJenergy}). The second subroutine takes into account the interactions between monomers located in different rings, see the second term in the right hand side of Eq.~(\ref{LJenergy}).
 Next, for each new conformation ${\cal X}^{(\nu+1)}$ that has been sampled, only the difference $\Delta E_{LJ}$ of energy between the new and the old conformations are computed.
 Let's suppose for instance that a new conformation
 ${\cal X}^{(new)}$ has been obtained from the old conformation ${\cal X}^{(old)}$ by a random transformation involving the monomers \verb|A+1|,....\verb|A+NPOINTS| of the knotted ring $K$.
 Then the energy $ E_{LJ}^{new}$ of the new conformation
is given by:
 \begin{equation}
  E_{LJ}^{new}=E_{LJ}^{old}+ \Delta E_{LJ}\label{Enew}
 \end{equation}
 where
 \begin{eqnarray}
\Delta E_{LJ}&=&   \frac 12   \sum_{I,J=1\atop J\ne I\pm 1}^{NPOINTS}
V_{LJ}(\boldsymbol r_{K,A+I}^{new} - \boldsymbol r_{K,A+J}^{new})\nonumber\\
&+&\sum_{I=1}^{NPOINTS}\sum_{J=1 \atop J\notin \{A,A+1,\ldots,A+NPOINTS+1\}}^L
V_{LJ}(\boldsymbol r_{K,A+I}^{new} - \boldsymbol r_{K,J}^{new})\nonumber\\
&+&\sum_{I=1}^{NPOINTS}\sum_{K'=1\atop K'\ne K}^{NKNOTS}
\sum_{J=1}^L
V_{LJ}(\boldsymbol r_{K,A+I}^{new} - \boldsymbol r_{K',J}^{new})+(new\Longleftrightarrow old)\label{deltaEIJ}
 \end{eqnarray}
 Despite the fact that Eqs.~(\ref{Enew}) and (\ref{deltaEIJ}) look much more complicated than the simpler Eq.~(\ref{LJenergy}), they have the advantage that the time $\tau_\Delta$ necessary for this calculation scales as $\tau_\Delta\propto 2\cdot \left(NPOINTS^2+NPOINTS\cdot NKNOTS\cdot L\right)$, the factor 2 arising because the calculation must be performed for both old and new conformations. Due to the fact that the number of monomers $NPOINTS$ interested in a random transformation is usually small compared to the total number of monomers $L$, $\tau_\Delta$ grows only linearly with the growth of the numbers of monomers \verb|NKNOTS| and knotted rings \verb|L|. 
 The subroutines that  compute the energy difference $\Delta E_{LJ}$ according to formulas
 ~(\ref{Enew}) and (\ref{deltaEIJ}) are called \verb|DENERGY2| and \verb|DENERGYR|..

 To conclude this Subsection, we discuss the approach used to sample conformations at low temperatures. This requires to find low energy conformations starting from a seed conformation that is usually not equilibrated, a problem that becomes challenging in the case of polymer systems of large size. The reason is that, once the system
 arrives at a deep local minima by random sampling,  it is likely to get trapped there, a situation that it is common while sampling at low temperatures.
 A typical scenario is that the system, starting from an out-of-equilibrium seed conformation of a given shape, stabilizes into conformations of approximately the same shape.
 To minimize the energy, as it is required at very low temperatures,
 local regions with increasingly higher monomer densities appear in the system, in which
 the distance between the monomers is such that the attractive component of the Lennard-Jones potential prevails. However, the large scale structure of the conformations does not change.  In other words, polymers "freeze" and, without appropriate measures, further sampling sticks to conformations in which the memory of the original shape is kept for a long time. The strategy implemented in the code to avoid bottlenecks of this kind, is the following.
 First, the sampling is performed at an initial temperature \verb|TEMPI|$>$\verb|TEMPF| which is high enough to allow a fast equilibration of the system.
 In a real system, this would mean that the energy $k_B\cdot$\verb|TEMPI| related to thermal fluctuations is much stronger than the potential energy of the  interactions among the monomers.
 In the present context,
recalling that thermodynamic units have been chosen such that $k_B=1$,
 the corresponding requirement is that the temperature $T=$\verb|TEMPI| appearing in the acceptance probability of Eq.~(\ref{metropolis}) is larger than the temperatures for which the LJ potential is dominated by the attractive
 component $\propto -r^{-6}$. For our purposes,
it is sufficient to choose \verb|TEMPI| several times larger than the parameter \verb|EPS| that determines the depth of the potential well of the LJ potential.
For instance, if \verb|EPS|$=1$, then a possible choice is \verb|TEMPI|$\ge 6.0$.
Successively, when the equilibrium is  attained at the temperature \verb|TEMPI|, the system is  "cooled" down gradually until the desired final temperature \verb|TEMPF| is reached. Concretely, if after a fixed number of sampled conformations (by default in the code this number has been set to $1.2\cdot 10^6$) the system is roughly found at equilibrium, then the temperature is decreased from the current temperature $T_s$ to the next lower one $T_{s+1}<T_s$, $s=0,\ldots,N$.
Otherwise, the last found conformation is taken as the new seed and the sampling is restarted at the temperature $T_s$. Of course, $T_0=$\verb|TEMPI| and the final temperature, after $N$ steps, is $T_N=$\verb|TEMPF|. 
To decide if the system is approximately at equilibrium at every temperature
$T_s$, the strength of the fluctuations of the specific heat capacity $c_V$, named \verb|CVTOT| in the code, is considered. The motivation of this choice is that the specific heat capacity is
proportional to the energy fluctuations $c_V=\frac{(\langle (E-\langle E\rangle )^2\rangle }{T^2\cdot L\cdot NKNOTS}$, where $\langle...\rangle$ denotes average over the sampled conformations. For this reason, $c_V$ can grow fast, especially at the beginning of the simulation, when the estimation of the
average $\langle E\rangle$ is still rough, so that the energy fluctuations may be large,
This effect  is further enhanced when the initial seed is out of equilibrium and the temperature is low, so that the factor $T^{-2}$ appearing in the denominator of the above expression of $c_V$, is not small. As a consequence, 
 the fluctuations of $c_V$ provide a suitable criterion in order to determine if the system is at equilibrium or not. 
Following this criterion, in the code the system is considered at equilibrium if
the relative error 
on the calculated specific heat capacity is less than 10\%:
\begin{equation}
  \frac{\langle c_V^2 \rangle-\langle c_V\rangle^2}{c_V^2}\le 0.1\label{maincondition}
  \end{equation}
The practical implementation of this strategy in the code is as follows.
  Starting from the initial temperature \verb|TEMPI|, the final temperature \verb|TEMPF|
  is reached in a number of \verb|TI=20| steps.
  Two nested loops in the file \verb|main.f90| are responsible for the sampling.
  After a number \verb|N=100000| of samples has been generated by the innermost loop, the partial results of the simulation are provided. They include the partial averages of the specific heat capacity, the elongation of the 
  knot due to the applied forces, and the square gyration radius.  Next to these values the related variances (specific heat capacity) or standard deviations (elongation and square gyration radius) are reported.
  The expectation values of the observables
  obtained from the innermost loop
  are further averaged by repeating this loop a maximum number \verb|N2=100000| of times.
  If the temperature is bigger than the final temperature, the iterations of the outer loop
  are stopped after \verb|N*24| conformations have been sampled if condition (\ref{maincondition}) is satisfied. In this case,  the temperature \verb|TEMP| is lowered by a quantity \verb|(TEMPI-TEMPF)/20| and the next \verb|N*24| samples are generated.
  If condition (\ref{maincondition}) is not verified, the last stored conformation becomes the new conformation, the temperature
  \verb|TEMP| remains unchanged and
  the procedure is repeated.
  In all the studied cases,  \verb|N*24| has proved to be the optimal frequency for checking the achievement of the requirement (\ref{maincondition}). If necessary, this frequency may be changed 
to any other multiple of \verb|N| by modifying the following line
  in \verb|main.f90|:
  \verb|IF((J2*1.0)/(24*1.0).EQ.J2/24) THEN|.
  When the final temperature is attained,
  the sampling is stopped after a number of \verb|N2*N|$=10^{10}$ conformations has been explored. For comparison, an additional set
of $=10^{10}$ conformations is generated in the last step \verb|TI|$=21$
and the expectation values of the observables are computed once again  at temperature \verb|TEMPF|.
\subsection{Measured observables}
The number of samples over which the expectation values of the desired observables are computed can be adjusted by changing the two integers
\verb|N2| and \verb|N|. In order to avoid large integer numbers, the summation over the samples has been splitted into two loops. The inner loop computes the 
 following expectation values averaged over \verb|N| samples:
\begin{enumerate}
\item  mean energy \verb|EAVE2|;
\item mean square energy \verb|ESAVE2|;
\item mean gyration radius \verb|RG|;
\item mean square gyration radius \verb|SRG|;
\item mean elongation of the system due to the force \verb|FORCEZ|. This measures the average absolute value of the distance between monomer no. \verb|PAF| of knot \verb|KNNZ| and monomer no. \verb|1| of knot \verb|1|.
\item mean-square elongation of the system
  \end{enumerate}
  The outer loop computes the final values of the quantities listed above.
  The total number of iterations is \verb|N2|$\cdot$\verb|N|. By default this number is set to $10^{10}$.
  For convenience, the partial values of the relevant quantities are provided after the end of the innermost loop, i. e. after \verb|J2*N| iterations, where \verb|J2=1,2,...| counts the number of iterations performed by the outermost loop and \verb|N| is the number of iterations in the innermost loop:
  \begin{enumerate}
  \item energy \verb |EN| of the last sampled conformation;
\item  mean energy $\langle E\rangle$  (\verb|<E>| in the output);
\item variance of the energy  $\langle (E-\langle E\rangle)^2 \rangle $ (\verb|<(E-<E>)^2>| in the output);
\item value of \verb|J2|;
\item partial specific heat capacity obtained after averaging over the samples generated during the last $N$ iterations of the innermost loop (\verb|CV-TEMP| in the output);
\item partial specific heat capacity averaged over \verb|J2*N| iterations \verb|CV| in the output);
\item variance of the specific heat capacity (\verb|ERR-CV| in the output);
  \item current value of the temperature (\verb|TEMP| in the output);
\item mean square gyration radius (\verb|RG| in the output);
\item standard deviation of the gyration radius (\verb|SRG| in the output);
\item current temperature (\verb|TEMP| in the output);
  \item value of the applied force (\verb|FORCE| in the output);
\item mean elongation of the system due to the applied force (\verb|ELONGATION| in the output);
\item standard deviation of the elongation (\verb|SELONG| in the output).
  \end{enumerate}
  The expectation value of any observable related to the conformations of the $NKNOTS$ rings can be added in the main program file \verb|main.f90|.
Finally, the last conformation sampled after a number \verb|24*N| conformations is stored in file \verb|polymer|. The last sampled conformation is saved in file \verb|polymer-final|.
  
  \section{Case studies}\label{apps}
  The first case study that will be discussed is that of a  polymer ring with $L=200$ monomers knotted in the topological configuration of the knot $4_1$.
  The settings for this series of runs are: \verb|TEMPI=3.0| (initial temperature), \verb|TEMPF=1.0| and \verb|TEMPF=0.5| (final temperatures), \verb|L=200| (total polymer length), \verb|NKNOTS=1| (number of knots), \verb|KNNZ=1| and \verb|PAF=L/2| (the force is applied to monomer  $L/2$ of knot 1).
  Several runs have been performed with  the force \verb|FORCEZ| along the $z-$axis taking the following different values:\\ $0.0,2.5,5.5,6.25,7.5,10,12.5,15.0,20,0,25,0, 30.0,35.0,40.0$.\\
  The fortran code was compiled with the GNU gfortran-15. A particularly convenient set of options for the compiler is: \verb|-Ofast --floop-nest-optimize -flto=auto|.
  It is also convenient to add the architecture related option \verb|-march=architecture|, where \verb|architecture| depends on the computer where the code is run. For example, on an AMD Ryzen Threadripper PRO 5995WX, \verb|-march=znver3|.
  On a Intel Core i7-12700K processor \verb|-march=alderlake|.
  Finally, in file \verb|nrotate.f90| the line:\\
  \verb|CALL TOP_CHECKR(REPNX,REPNY,REPNZ,KNN1,NKNOTS,A,FLAG1,LS)|\\
  should be commented when there is only a single knot.

\begin{figure}[h]
  \begin{center}
\includegraphics[width=0.48\textwidth]{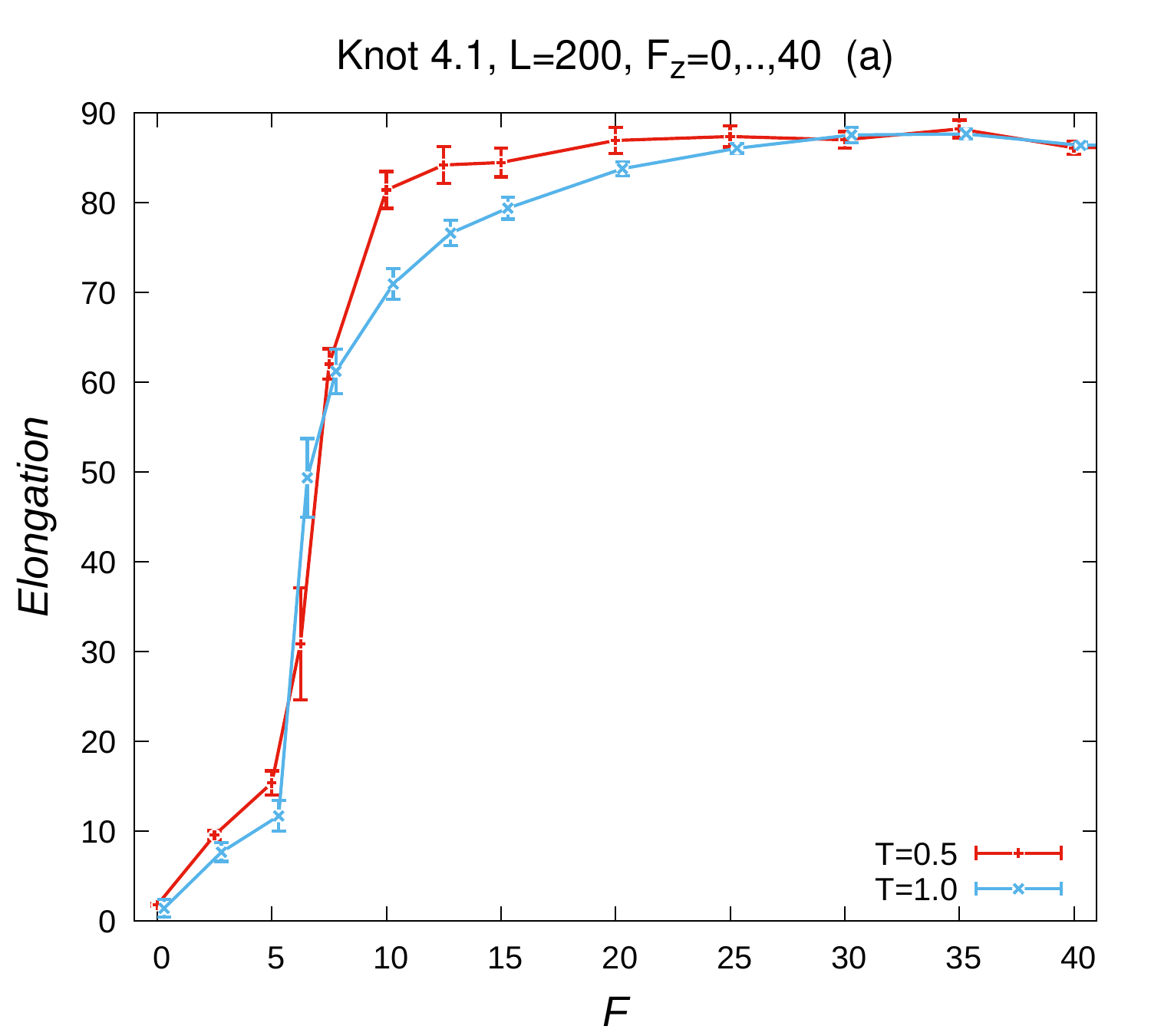}
\includegraphics[width=0.48\textwidth]{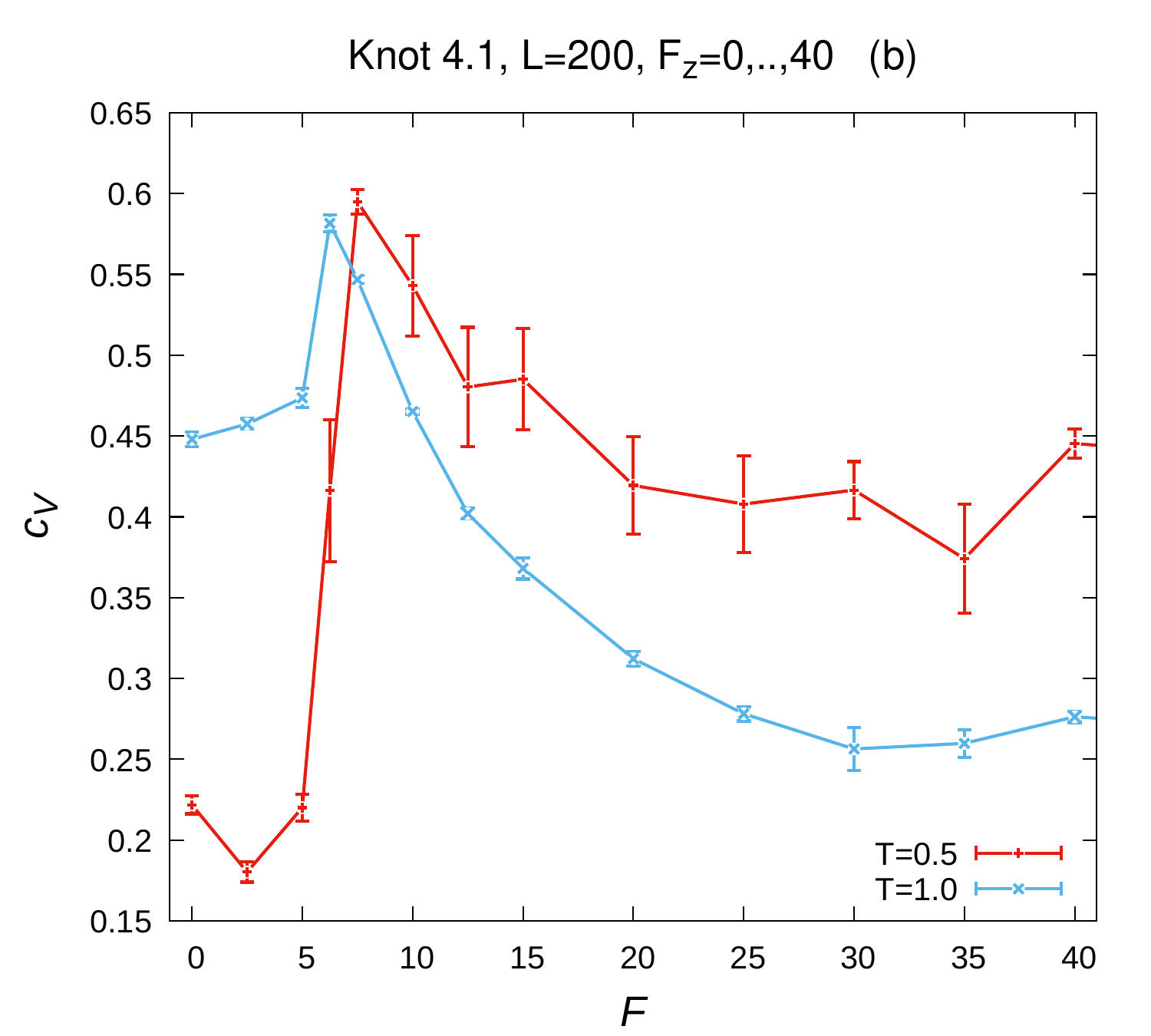 }
  \end{center}
  \caption
{Case of a single knotted polymer ring with $L=200$ and topology of the knot $4.1$. Panels (a) on the left and (b) on the right show respectively the force-elongation curves and the specific heat capacity at temperatures $T=0.5,1.0$.}
  \label{N200-phase-transition-colormaps1}
\end{figure}

The results of the simulations in the case of knot $4_1$ with $L=200$ are shown in Fig.~\ref{N200-phase-transition-colormaps1}. In panel (a) are shown the force-elongation curves, while in panel (b) the plot of the specific heat capacity $c_V$ is displayed.
Red color is used for the temperature $T=0.5$ and blue color in the case of $T=1.0$.
The steep growth of the elongation when the stretching force \verb|FORCEZ| (denoted $F$ in Fig.~\ref{N200-phase-transition-colormaps1}) is relatively small (between $5$ and $10$, see panel (a)), is compatible with the previous work \cite{YZFFPhysicaA2017} in which similar simulations have been performed on a simple cubic lattice.
The fact that the growth is smoother at the higher temperature $T=1.0$, as observed also in 
\cite{YZFFPhysicaA2017}, is due to the stronger thermal fluctuations that oppose the extension of the ring.
The plots of the specific heat capacity are in good qualitative agreement with those of Ref.~\cite{YZFFPhysicaA2017} too. In particular, even at high stretching forces, the values of the specific heat capacity remains almost constant and do not decrease. This is counterintuitive, because 
when the polymer is almost fully stretched, its entropy drastically decreases and
the conformations do not change considerably. As a consequence, the energy fluctuations that determine the specific heat capacity are not supposed to be large.
This apparent contradiction is solved by noting that a highly stretched knotted ring looks like a ladder, i. e. is long and narrow. The monomers located
on the two vertical rails of this ladder are very near to each other, so that even small fluctuations in their reciprocal distance can produce significant changes in the energy both in the case of the Lennard-Jones potential used here and of the short-range attractive interactions considered in Ref.~\cite{YZFFPhysicaA2017}.
At higher temperatures, for instance at $T=3.0$,
it has been checked that
the values of the specific heat capacity are almost zero, in agreement with Ref.~\cite{YZFFPhysicaA2017}.
The next  studied example is that of a poly[2]catenane consisting in an unknotted ring concatenated with an Hopf link  to a knotted polymer ring with the topology of the trefoil knot. Both rings have $L=24$ monomers.
 The settings in the code for this series of runs are similar to those of the previous case: \verb|TEMPI=3.0| (initial temperature), \verb|TEMPF=1.0| and \verb|TEMPF=0.5| (final temperatures), \verb|L=24| (total length of each ring), \verb|NKNOTS=2| (number of knots), \verb|KNNZ=2| and \verb|PAF=L/2| (the force is applied to monomer  $L/2$ of knot 2).
  As before,  the force \verb|FORCEZ| along the $z-$axis has been chosen to take the following different values:\\ $0.0,2.5,5.5,6.25,7.5,10,12.5,15.0,20,0,25,0, 30.0,35.0,40.0$.\\
  Of course, in file \verb|nrotate.f90| the line:\\
  \verb|CALL TOP_CHECKR(REPNX,REPNY,REPNZ,KNN1,NKNOTS,A,FLAG1,LS)|\\
  should not be commented.

\begin{figure}[h]
  \begin{center}
    \includegraphics[width=0.48\textwidth]{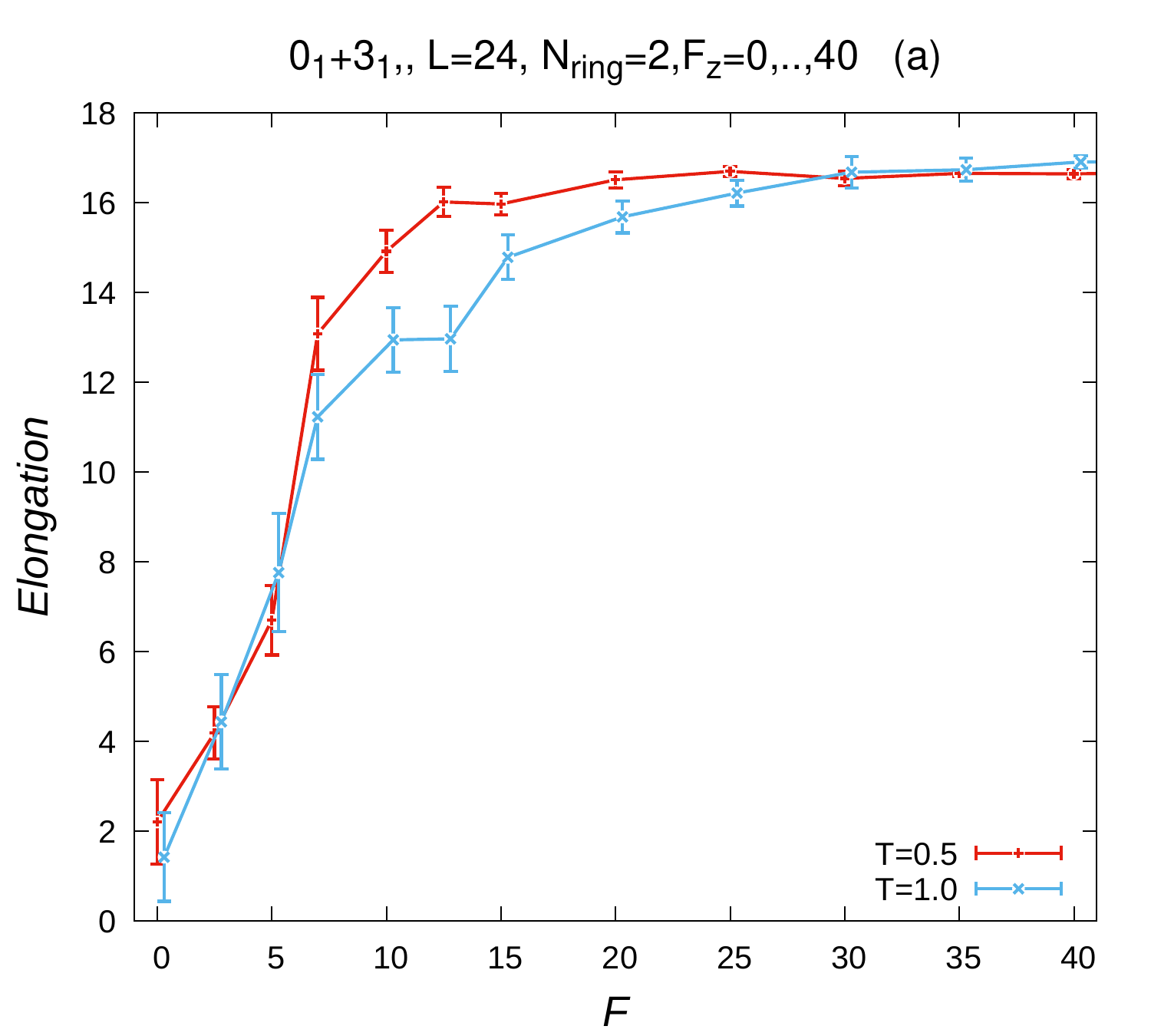}
    \includegraphics[width=0.48\textwidth]{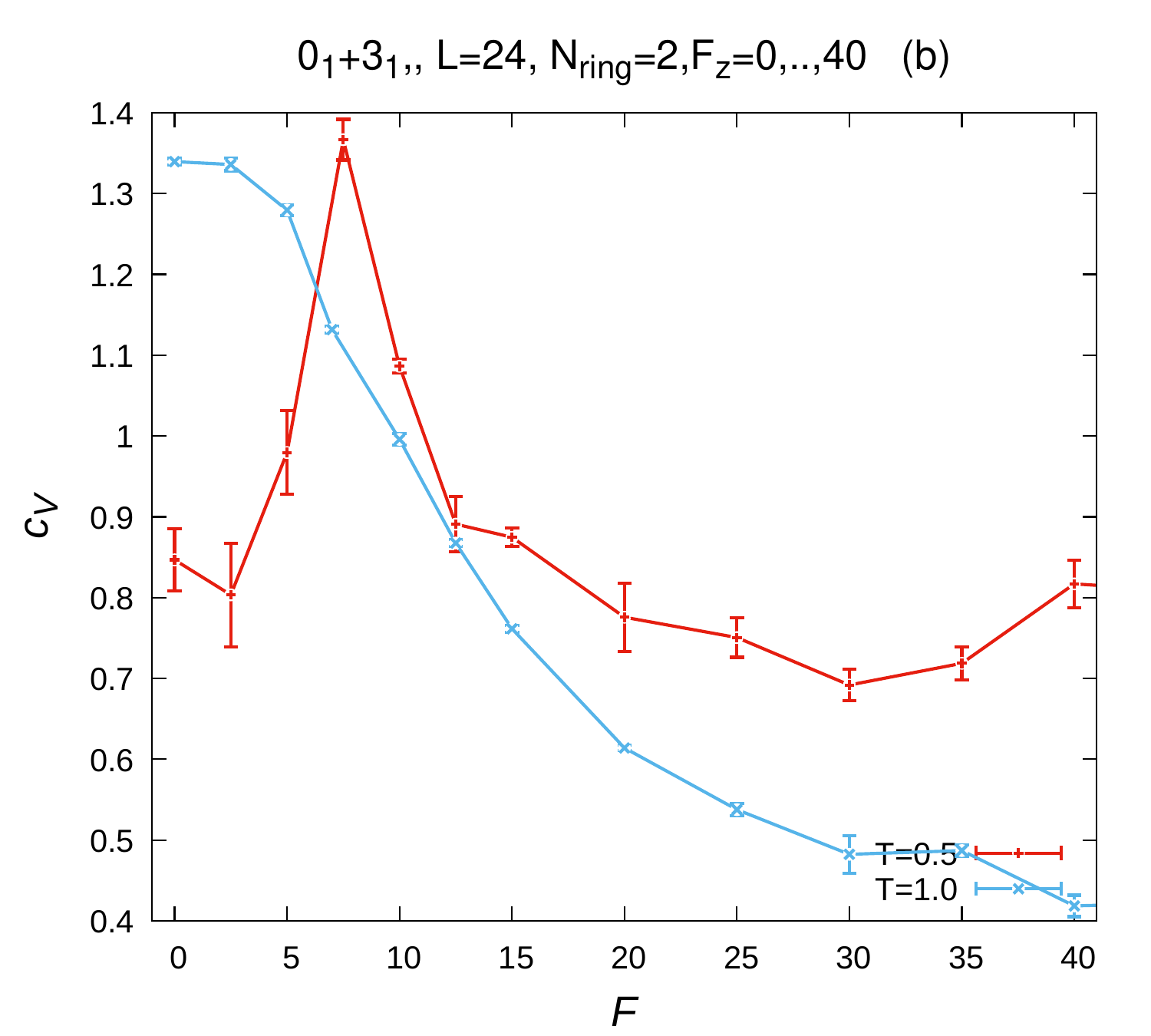}
  \end{center}
  \caption{Case of an unknot with $L=24$ concatenated with an Hopf link to a trefoil knot of the same length. Panels (a) on the left and (b) on the right show respectively the force-elongation curves and the specific heat capacity at temperatures $T=0.5,1.0$.} 
\label{N200-phase-transition-colormaps2}
\end{figure}

  The force-elongation curves of the poly[2]catenane at $TEMPF=0.5$ and
  $TEMPF=1.0$ are shown in Fig.~\ref{N200-phase-transition-colormaps2}, panel (a) and (b) respectively. We notice that, in all conformations that have been inspected,
   it has been observed that the knot $3_1$ is always strongly localized near the point in which the force is applied. 
  
\begin{figure}[h]
  \begin{center}
    \includegraphics[width=0.48\textwidth]{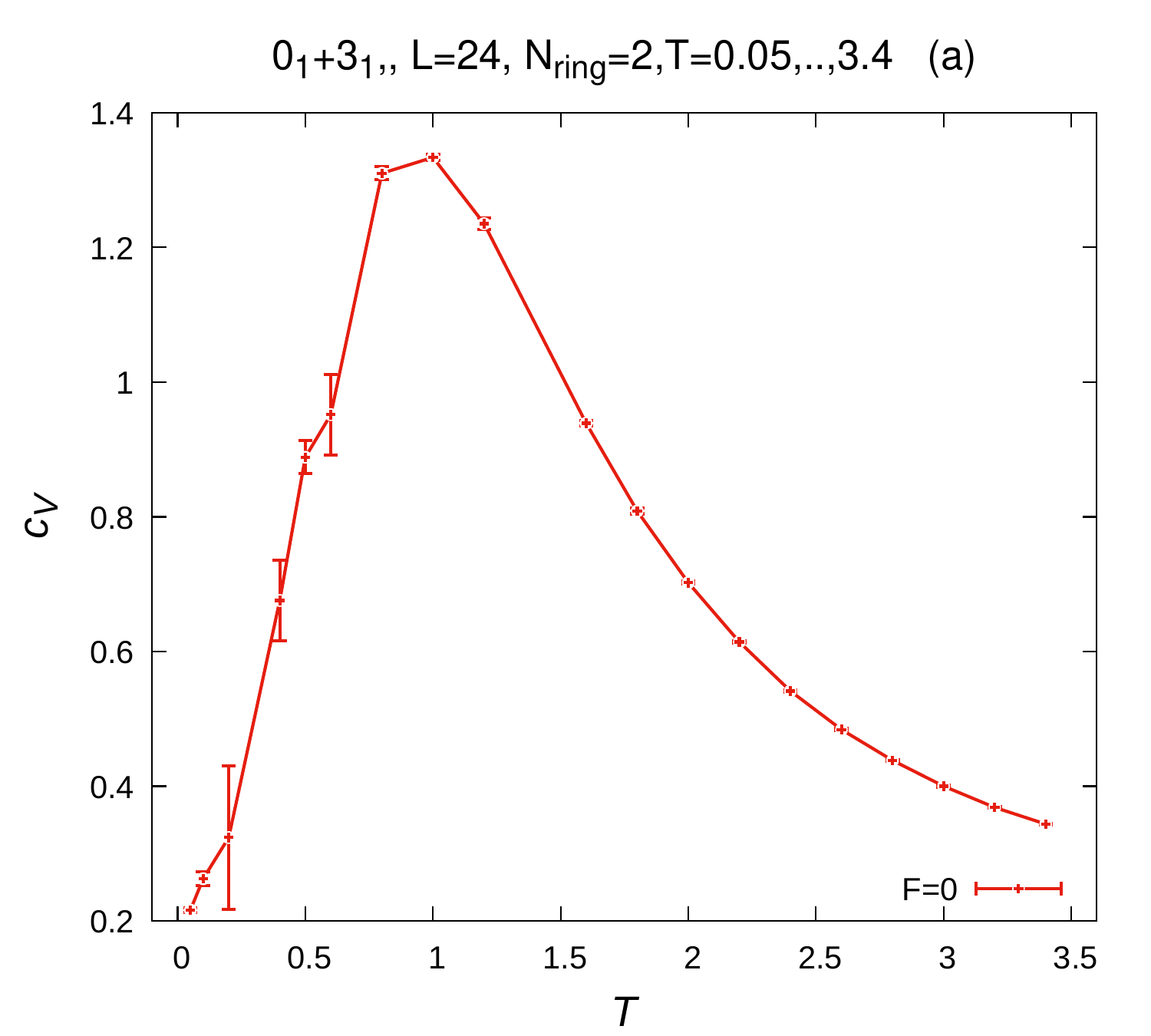}
    \includegraphics[width=0.48\textwidth]{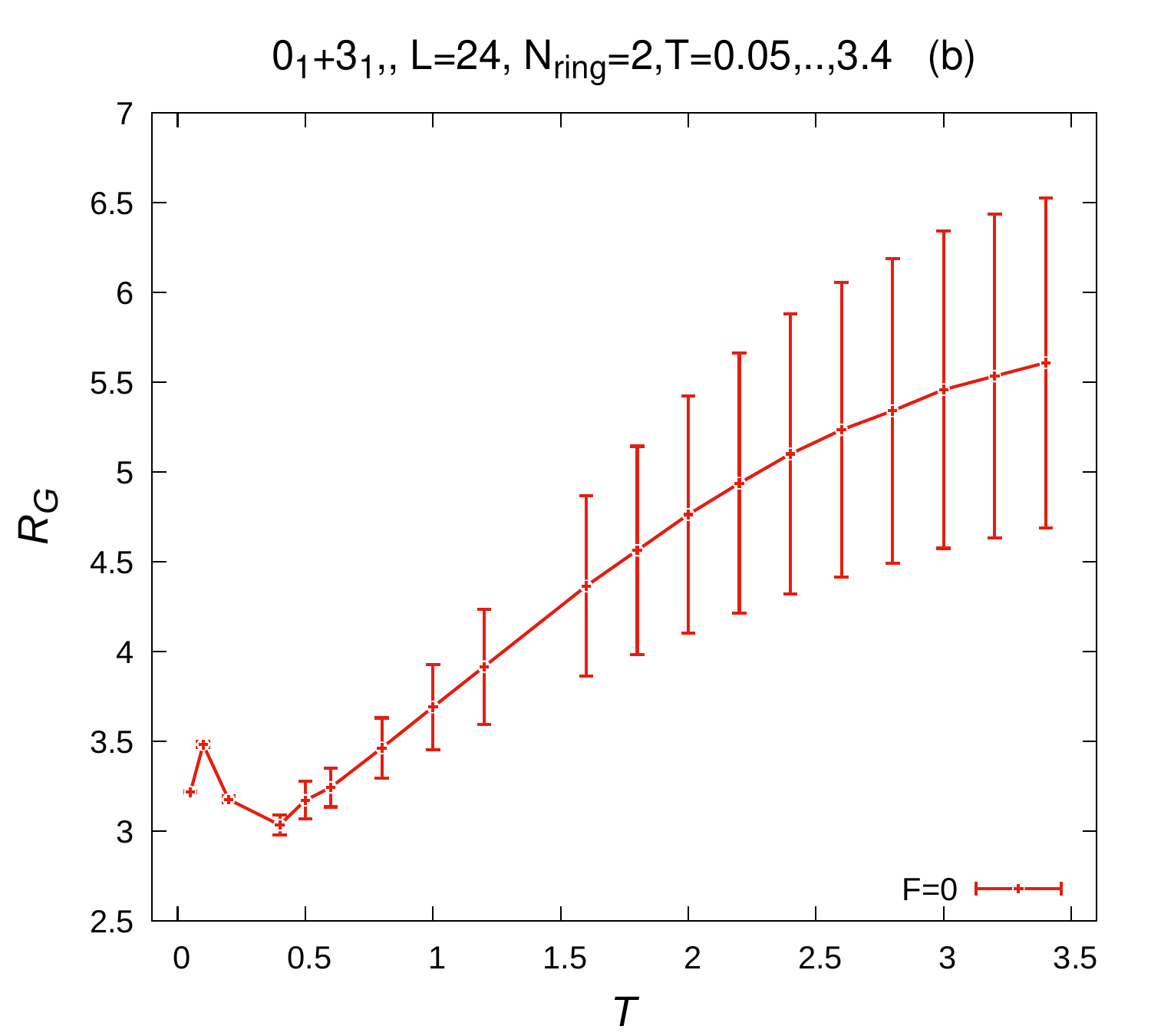}
  \end{center}
  \caption{Case of an unknot with $L=24$ concatenated with an Hopf link to a trefoil knot of the same length. A temperature sweep within the range $[0.05,3.4]$ was performed without applying any stretching force. 
    The results for the specific heat capacity and the gyration radius are shown in panels (a) and (b) respectively.}
\label{N200-phase-transition-colormaps3}
\end{figure}

\begin{figure}[h]
  \begin{center}
    \includegraphics[width=0.48\textwidth]{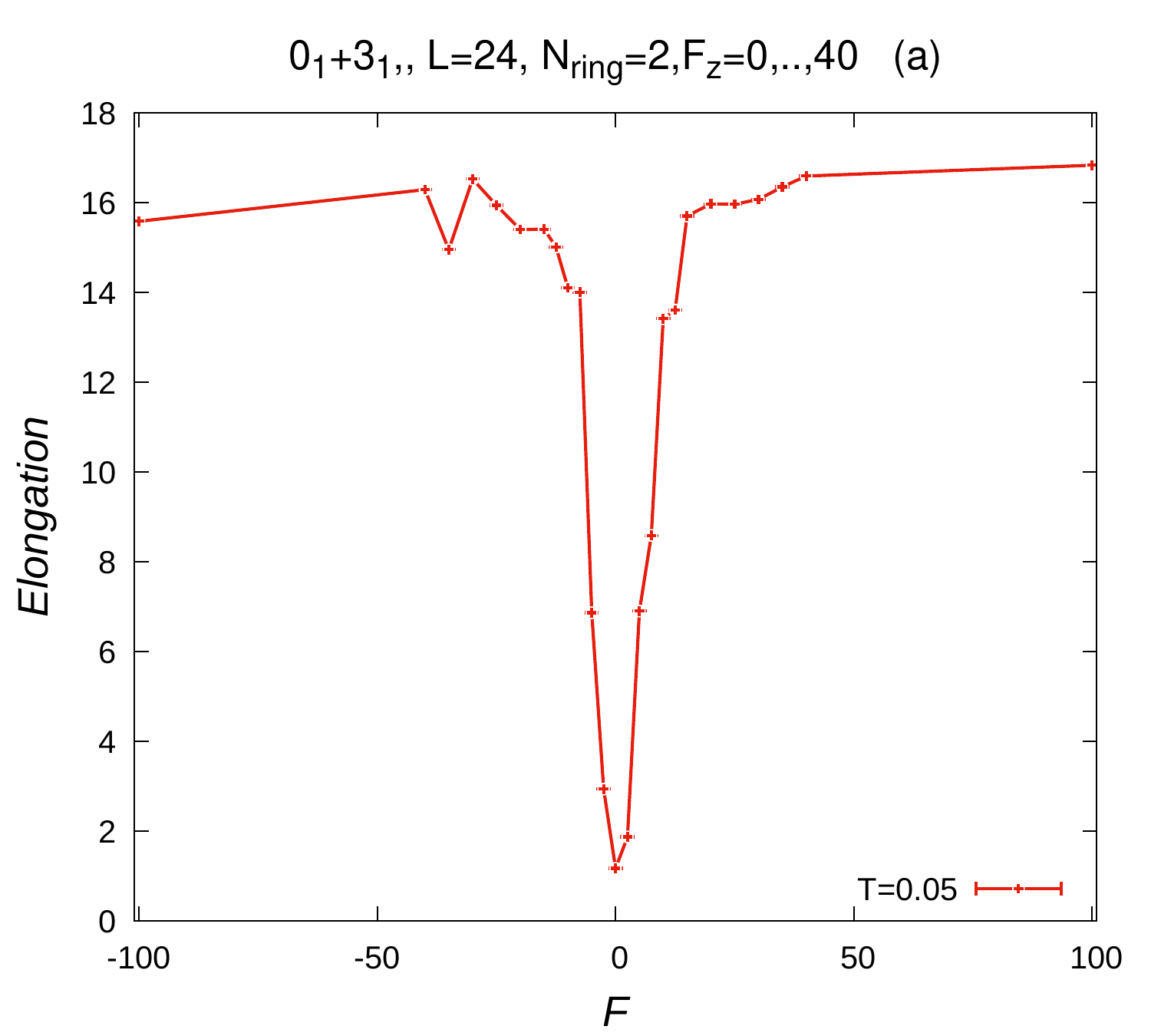}
    \includegraphics[width=0.48\textwidth]{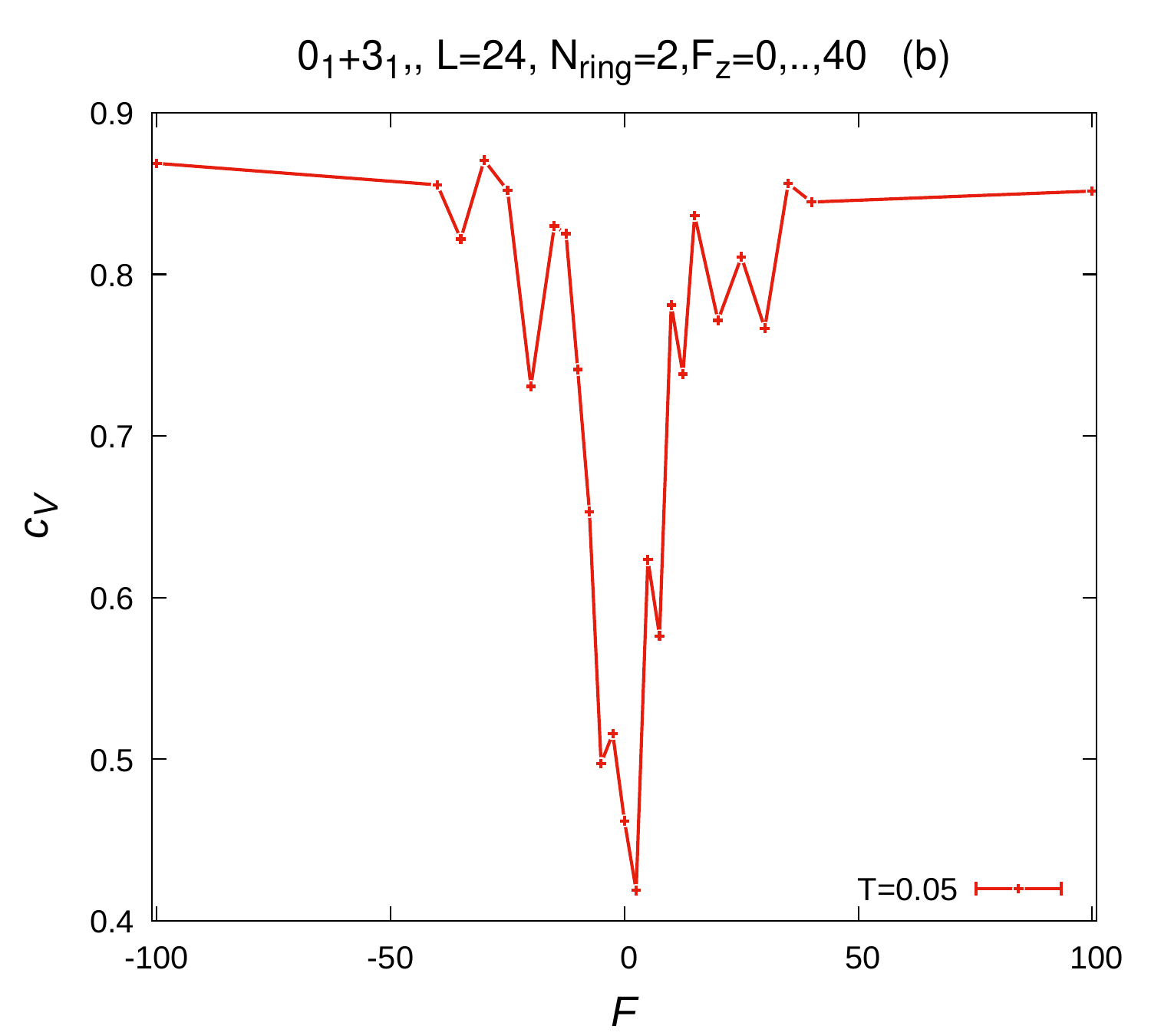}
  \end{center}
  \caption{Case of an unknot with $L=24$ concatenated with an Hopf link to a trefoil knot of the same length. Panels (a) on the left and (b) on the right show respectively the force-elongation curves and the specific heat capacity at the low temperature $T=0.05$. The range of the stretching force lies in the interval $[-100,100]$ } 
\label{N200-phase-transition-colormaps4}
\end{figure}

A temperature sweep within the range of temperatures $0.05\le T\le 3.4$ has been performed in the absence of any stretching force. The plots of the specific heat capacity and of the mean square gyration radius are presented in Fig.~\ref{N200-phase-transition-colormaps3}.
A possible asymmetry with respect to the sense of  \verb|FORCEZ| due to topology, for instance because $3_1$ is a chiral knot, has been checked by pulling the poly[2]catenane 
both up and down along the $z-$axis. Concretely, the system has been investigated in the case of the following values of \verb|FORCEZ|: $\pm 100,\pm 40,\pm 35,\pm 30,\pm 25, \pm20,\pm 15,\pm 12.5,\pm 10,\pm 7.5,\pm 6.25,\pm 5,\pm 2.5,0$.
The resulting force-elongation curve and the diagram of the specific heat capacity  are respectively shown in Fig.~\ref{N200-phase-transition-colormaps4}, panel (a) and panel (b).
As it is possible to notice, a slight asymmetry with respect to the sign of \verb|FORCEZ| is present.  Let's notice that the chosen temperature \verb|TEMPF=0.05| is very low. For this reason, the entropy of the system is minimal and the errorbars in the diagrams are negligible.

Finally, we discuss the simulations of a poly[32]catenane composed by $32$ unknots of length $L=40$ each. The total number of monomers is $1280$.
 The settings for this series of runs are: \verb|TEMPI=3.0| (initial temperature), \verb|TEMPF=1.0| and \verb|TEMPF=0.5| (final temperatures), \verb|L=40| (total length of each ring), \verb|NKNOTS=32| (number of rings), \verb|KNNZ=16| and \verb|PAF=L/2| (the force is applied to monomer  $L/2$ of ring $16$).
  The force \verb|FORCEZ| along the $z-$axis has been chosen to take the following different values:\\ $0.0,2.5,5.5,6.25,7.5,10,12.5,15.0,20,0,25,0, 30.0,35.0,40.0$.\\
  In file \verb|nrotate.f90| the line:\\
  \verb|CALL TOP_CHECKR(REPNX,REPNY,REPNZ,KNN1,NKNOTS,A,FLAG1,LS)|\\
  is not commented.
  In comparison with the previous, simpler polymers, the equilibration
  of the poly[32]catenane starting from an out of equilibrium seed is more complicated.
  In order to obtain equilibrated conformations the following strategy has proven to be very convenient.
 First, the system has been stretched at the relatively high temperature of \verb|TEMPF=5.0| using a very high pulling force: \verb|FORCEZ=400000|.
 A conformation of the almost fully elongated polymer obtained in this way (file \verb|polymer|) has been subjected to a smaller force \verb|FORCEZ=100| at temperature \verb|TEMPF=3.0|. Successively, the last conformation sampled after the simulation
with  \verb|FORCEZ=100| was stopped (file \verb|polymer-final|),
has been used as the starting seed of a new simulation with a lower force \verb|FORCEZ=40|.
  Continuing in this way up to \verb|FORCEZ=0|,
  It has been observed that the fluctuations in the specific heat capacity and other measured quantities stabilized fast, showing minimal variations\footnote{Longer polymers or polymers in which the rings are knotted to form complicated topologies, may require more efforts, like for instance repeating a few times the process of restarting a run
  by using the last stored conformation as the seed for the next run.
  It could also be necessary to perform a few passages at  higher temperatures  before reaching gradually the desired final temperature.
  }.
  The runs have been repeated a few times to verify the precision and accuracy of the calculations.
  
\begin{figure}[h]
  \begin{center}
    \includegraphics[width=0.48\textwidth]{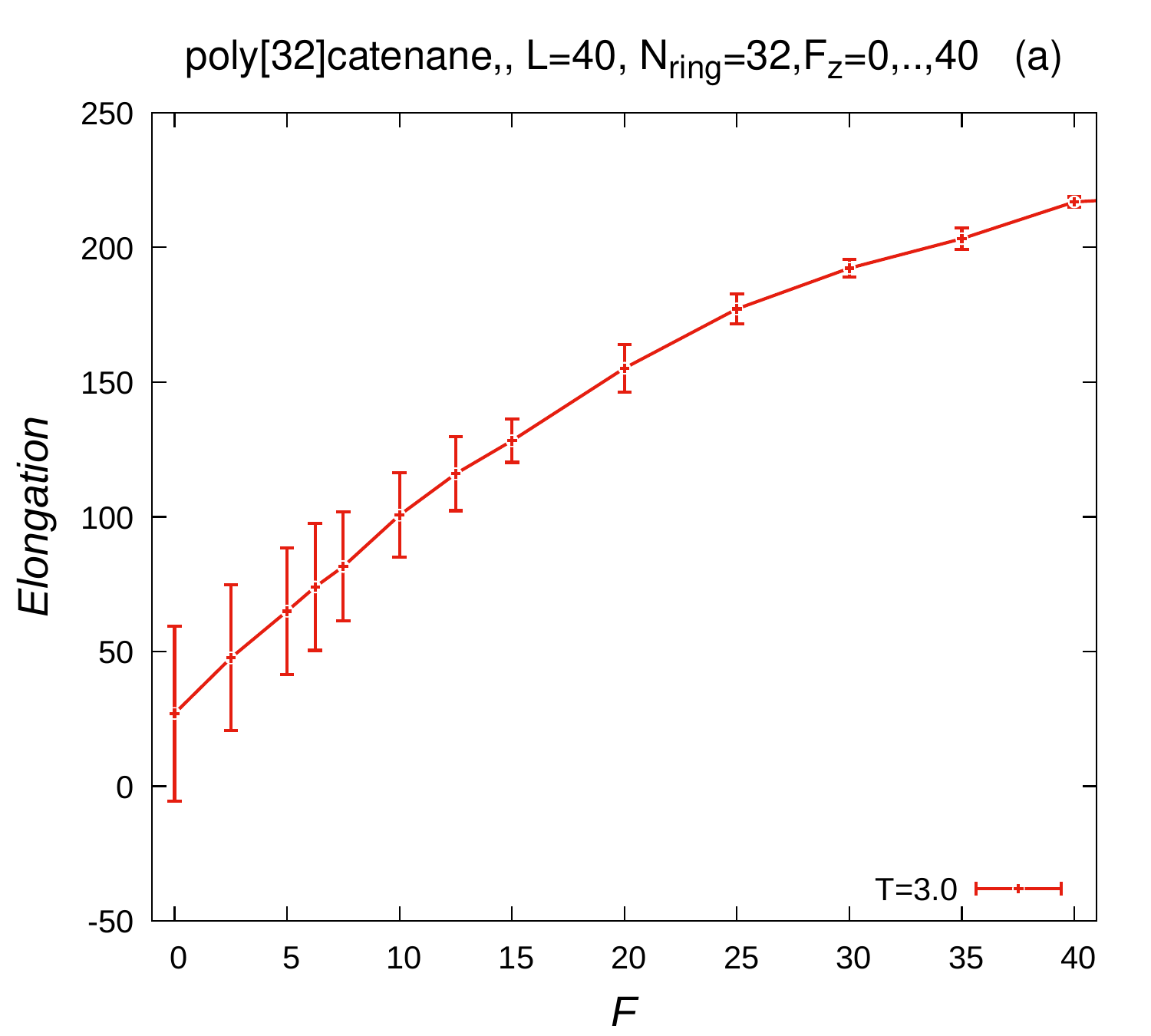}
    \includegraphics[width=0.48\textwidth]{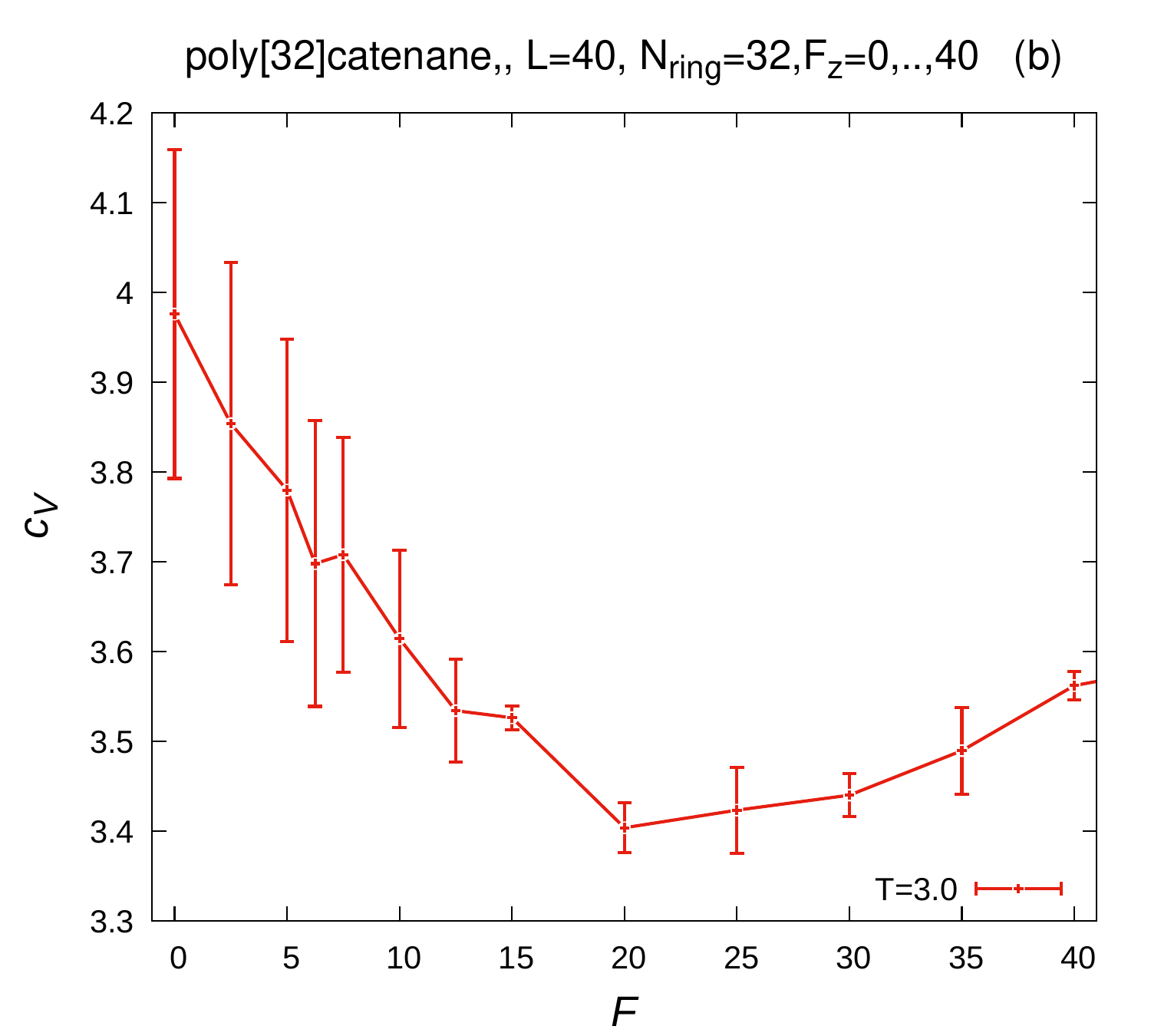}
  \end{center}
  \caption{Case of a poly[32]catenane composed by $32$ unknotted rings of  $L=40$
    monomers each concatenated together with Hopf links. Panels (a) on the left and (b) on the right show respectively the force-elongation curves and the specific heat capacity at temperatures $T=0.5,1.0$.} 
\label{N200-phase-transition-colormaps5}
\end{figure}

The force-elongation curves and the plot of the specific heat capacity for the poly[32]catenane are presented
in Fig.~\ref{N200-phase-transition-colormaps5}.

\section{Conclusions}\label{conclusions}
In this paper  the main features and the instructions have been presented of how to use SPOCK, a Monte Carlo code dedicated to the study of mechanical and thermal properties of knotted and/or concatenated polymer rings. The interactions are described by a Lennard-Jones potential and it is possible to apply to an arbitraryn number of monomers stretching forces. The topological constraints are preserved thanks to  a simple algorithm requiring that no part of the system
is inside the
region of space interested by a given random transformation if it is not directly involved in that transformation.

The simulations are fast also when polymers are relatively long.
For instance, all runs related to  the longest polymer studied here, namely the poly[32]catenane, ended up within less than six days on a AMD Ryzen 9950XCD processor.
The program has been tested so far with polymers containing up to 1980 monomers.
For the sake of speed gain, the algorithms used have been written in such a way that the time necessary to finish a simulation  scales up proportionally to the total number of monomers.
In the case of very long macromolecules, where parallelization could be helpful, the code can be easily parallelized using Coarray Fortran (CAF) and OpenMPI.
A list of future improvements in chronological order is: implementation of chain rigidity, addition of the interactions by a screened Coulomb potential, possibility of monomers of different type, and implementation of polymer chain elasticity.

The code will be made publicly available soon.

\section{Acknowledgements}\label{acknowledgments} 
The simulations reported in this work were performed in part using the HPC
cluster HAL9000 of
the University of Szczecin.
The research presented here has been supported by the Polish National Science Centre under
grant no. 2020/37/B/ST3/01471.
This work results within the collaboration of the COST
Action CA17139 (EUTOPIA).
The use of some of the facilities of the Laboratory of
Polymer Physics of the University of Szczecin, financed by 
a grant of the European Regional Development Fund in the frame of the
project eLBRUS (contract no. WND-RPZP.01.02.02-32-002/10), is
gratefully acknowledged.
F. F. is grateful to Michael Lang for the suggestion of investigating the poly[2]catenane composed by a $3_1$ knotted polymer linked with an unknot.

%% References
%%
%% Following citation commands can be used in the body text:
%% Usage of \cite is as follows:
%%   \cite{key}         ==>>  [#]
%%   \cite[chap. 2]{key} ==>> [#, chap. 2]
%%

%% References with bibTeX database:

\bibliographystyle{elsarticle-num}
%\bibliography{<your-bib-database>}

\begin{thebibliography}{00}
\bibitem{lammps} A. P. Thompson, H. M. Aktulga, R. Berger, D. S. Bolintineanu, W. M. Brown, P. S. Crozier, P. J. in 't Veld, A. Kohlmeyer, S. G. Moore, T. D. Nguyen, R. Shan, M. J. Stevens, J. Tranchida, C. Trott, S. J. Plimpton, {\it LAMMPS - a flexible simulation tool for particle-based materials modeling at the atomic, meso, and continuum scales},  {\it Comp Phys Comm} {\bf 271} (2022), 10817.
  \bibitem{gromacs} S. P\'all, A. Zhmurov,
P. Bauer,
M. Abraham,
M. Lundborg,
A. Gray,
B. Hess,
E. Lindahl, {\it Heterogeneous parallelization and acceleration of molecular dynamics simulations in GROMACS}, {\it J.  Chem. Phys.} {\bf 153} (2020), 134110.
\bibitem{hoomd} J. A. Anderson, J. Glaser, S. C. Glotzer,
{\it HOOMD-blue: A Python package for high-performance molecular dynamics and hard particle Monte Carlo simulations},
{\it Computational Materials Science} {\bf 173} (2020),
109363,
\bibitem{dlpoly} W. Smith and T. R. Forester, 
  {\it  A general-purpose parallel molecular dynamics simulation package},
  {\it Jour. of Molecular Graphics} {\bf 14}(2) (1996), 136-136.
\bibitem{baumgartner} A. Baumg\"artner, {\it Statics and dynamics of the freely jointed
polymer chain with Lennard-Jones
interaction}, {\it J. Chem. Phys.} {\bf 72} (1980), 871.
  \bibitem{YZFFPhysicaA2017} Y. Zhao and F. Ferrari, {\it Physica A: Statistical Mechanics and its Applications},
    {\bf 486} (2017), 44-64.

   
%% \bibitem must have the following form:
%%   \bibitem{key}...
%%

% \bibitem{}

 \end{thebibliography}

%% Authors are advised to submit their bibtex database files. They are
%% requested to list a bibtex style file in the manuscript if they do
%% not want to use elsarticle-num.bst.

%% References without bibTeX database:

\end{document}